\documentclass[a4paper,12pt]{article}
\usepackage[T1]{fontenc}
\usepackage{fullpage}
\usepackage{graphicx}
\usepackage{amsmath}
\usepackage{color}
\usepackage{cancel}
\usepackage[normalem]{ulem}
\usepackage[left]{lineno}
\usepackage{blindtext}
\usepackage{subfig}
\usepackage{hyperref}
\title{Scaling laws in global corporations as a benchmarking approach to assess environmental performance}
\author{\small{\textit{Rossana Mastrandrea$^{1\ast}$ ,Rob ter Burg$^{2\ast}$, Yuli Shan$^{3}$, Klaus Hubacek $^{2}$ \& Franco Ruzzenenti$^{2\star}$}}}
\date{}

\begin{document}
	
\maketitle
\begin{small}
\begin{center}
\textit{
$^{1}$ IMT School for Advanced Studies, Lucca, Piazza S. Ponziano 6, 55100 Lucca, Italy\\
$^{2}$ Integrated Research on Energy, Environment and Society (IREES), Faculty of of Science and Engineering, University of Groningen, Groningen, 9747 AG, the Netherlands.\\
$^{3}$ School of Geography, Earth and Environmental Sciences, University of Birmingham, Birmingham B15 2TT, UK.\\
}
\end{center}

\noindent
\begin{footnotesize}$^{\ast}$ These authoirs equally contributed to the work. $^{\star}$ Corresponding author:  f.ruzzenenti@rug.nl, \end{footnotesize}\\

\medskip

\noindent
Keywords: Allometry; Transnational Corporations; Corporate Metabolism; Environmental and Social Governance; Emission Targets; Zipf's law; Strategic Benchmarking.
\end{small}

 
\begin{abstract}
The largest 6,529 international corporations are accountable for almost 30\% of global $\text{CO}_{2}e$ emissions. A growing awareness of the role of the corporate world in the path toward sustainability has led many shareholders and stakeholders to pursue increasingly stringent and ambitious environmental goals. However, how to assess the corporate environmental performance objectively and efficiently remains an open question. This study reveals underlying dynamics and structures that can be used to construct a unified quantitative picture of the environmental impact of companies. This study shows that the metabolism of global corporations in terms of: $\text{CO}_{2}$, energy used, water withdrawal and waste production, scales with their size according to a simple power law which is often sublinear, and can be used to derive a sector-specific, size-dependent benchmark to asses unambiguously a company's environmental performance. Enforcing such a benchmark would potentially result in a 15\% emissions reduction, but a  fair and effective environmental policy should consider the size of the corporation and the super or sublinear nature of the scaling relationship.

\end{abstract}

In 2018 less than seven thousand of the largest international corporations, accounting for 50\% of World GDP and employing 123 million people, released more than 14 billion tonnes of $\text{CO}_{2}$ equivalent, i.e. 30\% of global emissions, with an average intensity of 117 tonnes per worker, three times higher than the most densely populated mega-cities \cite{wiedmann2021three}. 
ESG (Environmental, Social Governance) practices and policies are meant to address emission reduction targets. However, many voices, corporate and non-corporate, are being raised over the issue of the lack of a transparent and unbiased \emph{benchmark} to assess the environmental performance of companies \cite{kishan_bloomberg,carbon_market_watch2022}. We propose here a bench-marking approach based on scaling analysis of corporate metabolism and self-reporting (EIKON database) that could represent an easy solution to gauge how a company performs compared to the sector to which it belongs and its size, according to the most suitable metrics (revenue, employees or capitalization).  

Ahead of the UN Climate Action Summit  of 2019, a coalition of almost 90 major global corporations committed to bring their emissions to zero before 2050 \cite{reuters_big_companies}. This is the last step in a process of progressive engagement by the corporate world into a decades' long path toward sustainability, which began in the aftermath of the Rio Earth Summit of 1992 \cite{elkington1994towards} and the foundation of the World Business Council for Sustainable Development\cite{najam2013world}. This commitment has become more stringent since it was joined by the sphere of finance, in a parallel quest for sustainability which began essentially 20 years ago with the Equator Principles and culminated in 2019 with the Principles for Responsible Banking \cite{griffiths2021or}. All these initiatives have been so far voluntary in nature and have led to very different ESG practices and approaches, with different results and degree of alignment with the proclaimed environmental goals \cite{ihlen2009business,ameer2012sustainability} or emission targets \cite{dietz2018assessment}; on the other hand, this commitment was met with a growing and vocal call for greater (biosphere) stewardship by transnational corporations from science and other societal actors \cite{folke2019transnational}.
This willingness to engage, however, come with some major challenges, the most cogent of which, is that of \emph{assessing the real impact} of corporate activity, both in terms of emissions and other environmental pressure \cite{crona2021transforming}. Global value chains make this task even more difficult \cite{zhang2020embodied,zhang2021using}, but the problem of establishing a harmonized analytical framework and unbiased metrics is long standing \cite{wiedmann2006triple,hoekstra2014humanity}.The issue of how to \emph{benchmark} corporate emissions and environmental impact, once the conundrum of scholars and the concern of policy makers and environmentalists, is now haunting  financial investors (institutional and not) who are seeking the compliance with the forthcoming stringent rules of series of compulsory legal and regulatory interventions, such as the Sustainable Finance Action Plane of the European Union \cite{EUsusfin}. Typically, existing ratings have a lack of scientific basis and cross-country and cross-sector comparability\cite{burnside2019corporations}. Despite few attempts, and at least until the zero-emission target will be achieved, the question of how to gauge objectively the emissions of a corporation remains unresolved \cite{karim2021novel,wedari2021corporate}. At the present time, there is no general, universally accepted criteria upon which the environmental impact of companies is measured \cite{dahlsrud2008,vigneau2015firms,kpmg2017}. Moreover, the measurement methods that do exist are typically prone to subjectivity as they are based on self-reporting \cite{galant2017}. Several attempts have been made in order to standardize sustainability reporting, such as the GRI Sustainability Reporting Standards \cite{GRI}and the United Nations Principles for Responsible Investments \cite{pri}. In the supplementary material (SM1) we offer a brief overview of these approaches, pros and cons \cite{friedman2007,kpmg2017,ioannou2017consequences} and the methodologies used to quantify them \cite{galant2017,huber2017}. Parallel to the problem of assessing environmental performance, the scientific and corporate community engaged in finding a scientific-based methodologies for setting emission targets in line with climate targets \cite{bjorn2021paris}; and among them those considering sector-specific emissions' reduction pathways, such as the Sectoral Decarbonization Approach, took into consideration the constrains peculiar of every sectors/industries in determining such paths\cite{krabbe2015aligning}. Although the business sector has been recognized as a factor in defining objectives and evaluating performance, the size of the company has not so far received an equivalent attention. Here, we will try to fill this gap by showing that environmental and emissions benchmarks should be adapted to the size of the company and the reduction targets should consider whether its sector scales sub(super)linearly, on logarithmic scale, as emission reductions would result in a lower (greater) burden for larger companies than for small businesses.   


\section*{Results}
The four Environmental Impact indicators selected from the EIKON database to asses corporate metabolism are: $\text{CO}_2$ emissions, energy use, water withdrawal or waste. For every dependent variable (= the environmental indicator) a linear regression model was fitted using firm size, i.e. assets, employees, market capitalisation and total revenue, as independent variables. In its linear form, the constants $\beta$ (slope) and $\eta$ (intercept) in the scaling equations can be assessed by performing a linear regression analysis for the 10 main sectors or 133 industries (see Methods).  The definition of industries and sectors follows the Refinitiv Business Classification (TRBC), developed by the Reuters Group under the name Reuters Business Sector Scheme \cite{refinitiv}. 
The two constants are specific for every sector (or industry) and determine unambiguously the environmental impact of a company according to the incumbent benchmark.  Applying globally this benchmark would potentially result in a 15\% emissions reduction. If more and more companies would perform better or as good as the benchmark, the benchmark would shift downward, enhancing the environmental standard, until some limiting environmental and technical constraints are reached.    
\subsection*{Emissions}
\begin{figure}[ht]
\begin{center}
\centerline{\includegraphics[width=1\textwidth]{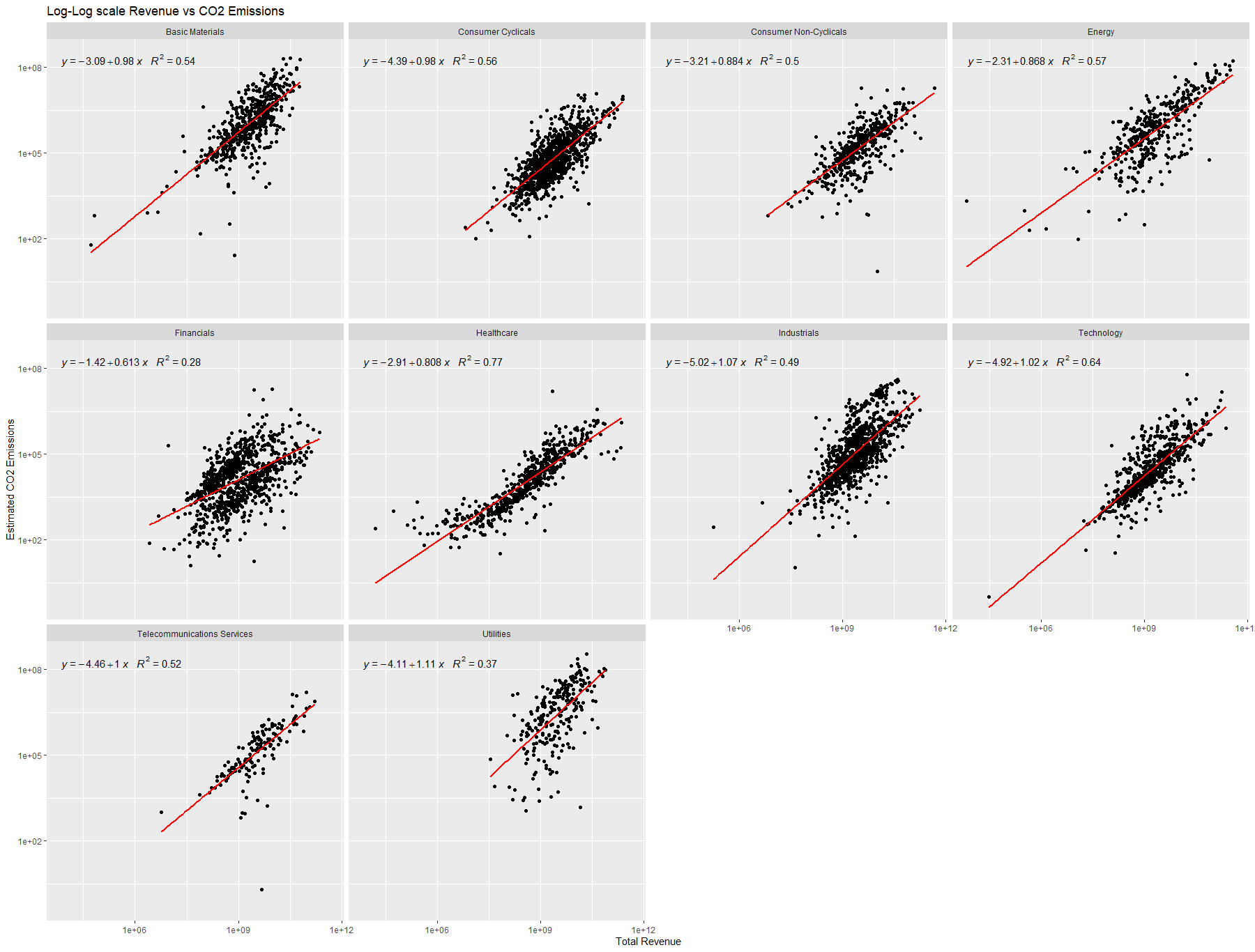}
}
\caption{\textbf{Log-log scale plot of total revenue vs. total emissions by sector (including the $\text{R}^2$-value). The scaling plot improves significantly in terms of fitting with the granularity of the data and best fitting it is found at the industry level (see supplementary data).}\label{fig_sectors}}
\end{center}
\end{figure}

When plotted on a log-log scale, total revenue and estimated $\text{CO}_2$ equivalent emissions for all the companies show a linear relationship (Figure \ref{fig3}), first quadrant) with $\text{R}^2$-value of 0.45, despite the variety in companies, industries and economic sectors. In the plot several clouds of sectors can be observed and by dis-aggregating the data the fitting improves: at sector level of aggregation (Figure \ref{fig_sectors}) the fitting shows a mean $\text{R}^2$ of 0.58; the best fitting is obtained at industry level (see supplementary information). The three most emission intensive economic sectors were found to be utilities, basic materials and energy. Companies from these sectors were found to often be above the regression line (i.e. benchmark). On the contrary, companies of sectors with lower emissions intensities lie often below the benchmark. Examples of such sectors are healthcare and finance. 
The slope of the regression equation shows a sublinear relationship, with $\beta$=0.94. Sublinearity suggests that the growth in emissions is bounded. Thus, as companies grow, they become more efficient in terms of emissions per € revenue (or other size unit). Most of industries show sublinearity, with few exceptions, like Airlines, Marin Freight, Power Equipment, Commodity Chemicals and Rubbers (see SM2 for a the full list).
With respect to the estimated equivalent emissions, the total revenue, number of employees and total assets seems to be good predictors. Of these size indicators, the total revenue was proved to have the highest correlation and best fit to the regression model. Most industries, 119 out of 123 respectively, showed to have a statistically significant relationship with emissions at the 5\% level and 107 industries at the 0.1\% level. From the size indicators, market capitalisation was on average the worst predictor. Merely slightly more than half of the industries showed a significant relationship at 0.1\% level. Additionally, market capitalisation also showed the lowest correlation and $\text{R}^2$-values (see Table \ref{tab:emissions}).

\begin{table}[]
\caption { $\text{CO}_2$ equivalents emission vs. company size indicator} \label{tab:emissions} 
\resizebox{\textwidth}{!}{%

\begin{tabular}{llllllllllllll}
                            &       & \multicolumn{3}{l}{Employees} & \multicolumn{3}{l}{Market Capitalisation} & \multicolumn{3}{l}{Assets}  & \multicolumn{3}{l}{Total Revenue} \\
Sector                      & n     & Adj. $\text{R}^2$  & $\beta$   & c      & Adj. $\text{R}^2$      & $\beta$       & c          & Adj. $\text{R}^2$ & $\beta$  & c      & Adj. $\text{R}^2$   & $\beta$    & c        \\
All                         & 6,529 & 0.334    & 0.799***  & 1.959  & 0.127        & 0.544***      & -0.215     & 0.337   & 0.918*** & -3.863 & 0.446     & 0.944***   & -3.798   \\
Basic Materials             & 635   & 0.499    & 0.978***  & 2.377  & 0.229        & 0.700***      & -0.497     & 0.520   & 1.097*** & -4.402 & 0.608     & 0.951***   & -2.798   \\
Consumer Cyclicals          & 1,034 & 0.633    & 0.979***  & 0.921  & 0.230        & 0.534***      & -0.154     & 0.508   & 0.940*** & -4.089 & 0.569     & 0.988***   & -4.458   \\
Consumer Non-Cyclicals      & 481   & 0.569    & 0.824***  & 1.900  & 0.264        & 0.538***      & 0.076      & 0.538   & 0.915*** & -3.516 & 0.545     & 0.878***   & -3.133   \\
Energy                      & 420   & 0.344    & 0.691***  & 3.505  & 0.373        & 0.689***      & -0.391     & 0.591   & 1.078*** & -4.556 & 0.582     & 0.850***   & -2.068   \\
Financials                  & 993   & 0.193    & 0.432***  & 2.720  & 0.195        & 0.614***      & -1.673     & 0.263   & 0.609*** & -1.939 & 0.272     & 0.613***   & -1.461   \\
Healthcare                  & 659   & 0.433    & 1.018***  & 0.714  & 0.856        & 0.823***      & -3.698     & 0.745   & 1.132*** & -6.214 & 0.792     & 0.810***   & -2.902   \\
Industrials                 & 1,137 & 0.375    & 0.868***  & 1.590  & 0.166        & 0.546***      & -0.060     & 0.473   & 1.013*** & -4.611 & 0.470     & 1.000***   & -4.331   \\
Technology                  & 742   & 0.570    & 0.990***  & 0.708  & 0.205        & 0.553***      & -0.910     & 0.555   & 0.994*** & -4.858 & 0.592     & 0.980***    & -4.543   \\
Telecommunications Services & 153   & 0.504    & 0.862***  & 1.742  & 0.268        & 0.638***      & -1.016     & 0.449   & 0.927*** & -4.073 & 0.505     & 0.946***   & -3.909   \\
Utilities                   & 275   & 0.238    & 0.829***  & 3.316  & 0.150        & 0.757***      & -0.950     & 0.359   & 1.231*** & -5.970 & 0.357     & 1.144***   & -4.488  
\\
* p\textless{}0.05, **p\textless{}0.01, ***p\textless{}0.001         
\end{tabular}%
}
\end{table}

\subsection*{Energy use}
\begin{figure}[ht]
\begin{center}
\centerline{\includegraphics[width=1\textwidth]{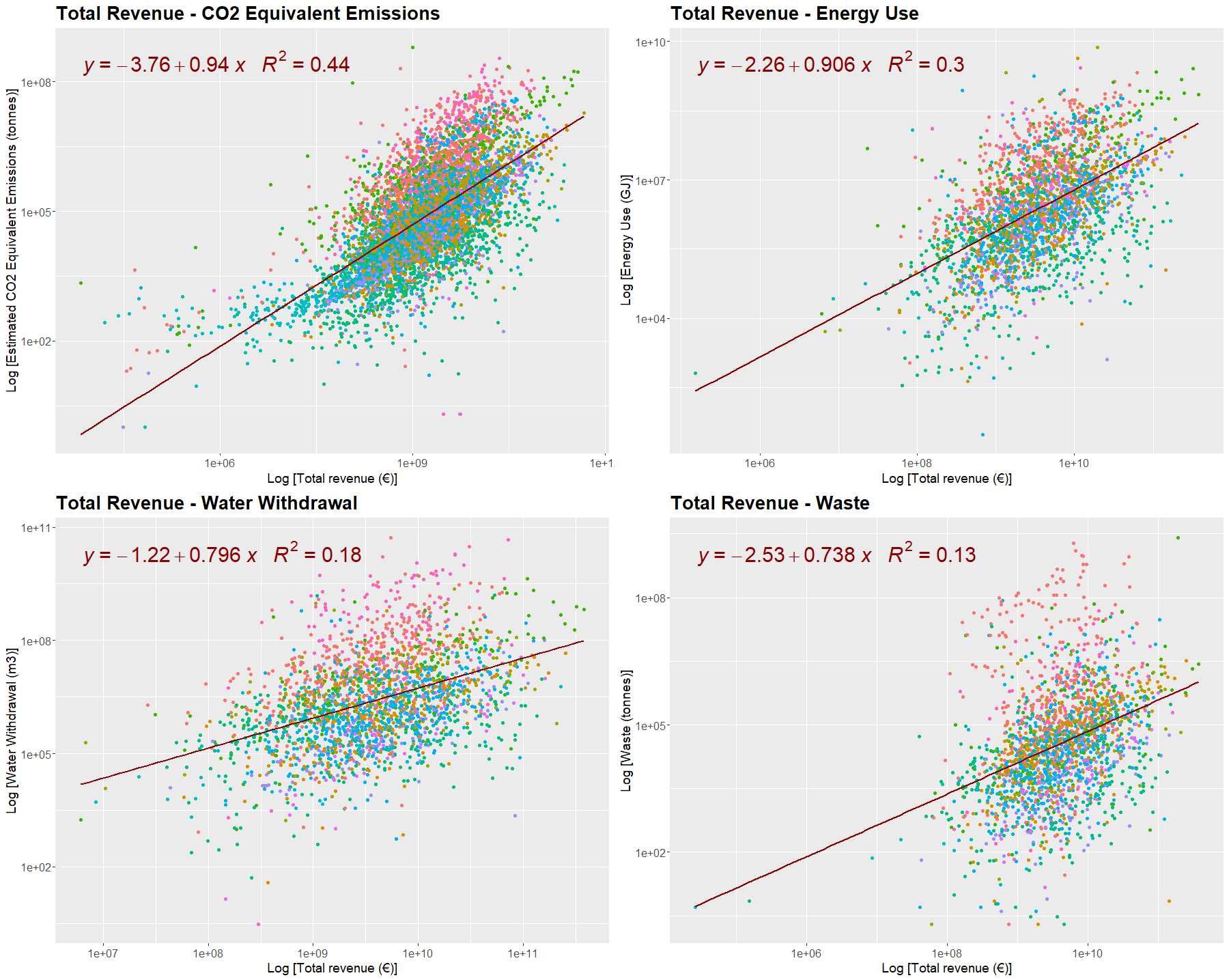}
}
\caption{\textbf{Log-log scale plot of total revenue vs. the four impact indicators: emissions, energy, water and waste, different colours indicate different economic sector. Red line = linear regression line, with the regression model depicted on the left top (including the $\text{R}^2$-value)}\label{fig3}}
\end{center}
\end{figure}

Information about the energy use of companies is not as widely available as the estimated $\text{CO}_2$ equivalent emissions. Therefore, only 2,416 companies were included in the sample for analysing the scaling of energy use. There is only a marginal difference in $\beta$, namely 0.92 for energy use compared to the 0.94 for the emissions (Figure \ref{fig3}).
Splitting the sample into industries again proved to give a better fit as the $\text{R}^2$ increased for every size variable. On average, revenue had the highest adjusted $\text{R}^2= 0.47$ closely followed by assets with an adjusted $\text{R}^2$ of 0.46. Moreover, employees also showed a good average adjusted $\text{R}^2=0.42$. The adjusted $\text{R}^2$-values of energy use are lower than for the emissions, implying that the $\text{CO}_2$ equivalent emission data fit the regression models better than the energy use data. In contrast to the exponents of the emissions, most of the industries scale superlinearly with revenue, as can be seen in Figure \ref{fig3}. More precisely, 43 of the 91 industries scale superlinearly, meaning that the relative increase in energy consumption is higher than the relative increase in size.

\begin{table}[]
\caption {Energy use vs. company size indicator} \label{tab:energy} 
\resizebox{\textwidth}{!}{%
\begin{tabular}{llllllllllllll}
                            &       & \multicolumn{3}{l}{Employees} & \multicolumn{3}{l}{Market Capitalisation} & \multicolumn{3}{l}{Assets}  & \multicolumn{3}{l}{Total Revenue} \\
Sector                     & n     & Adj. $\text{R}^2$  & $\beta$   & c      & Adj. $\text{R}^2$      & $\beta$       & c          & Adj. $\text{R}^2$ & $\beta$  & c      & Adj. $\text{R}^2$   & $\beta$    & c        \\
All                         & 2,416 & 0.213    & 0.687***  & 3.687  & 0.074        & 0.437***      & 2.252      & 0.189   & 0.744*** & -0.911 & 0.299     & 0.919***   & -2.371   \\
Basic Materials             & 373   & 0.421    & 0.966***  & 3.531  & 0.136        & 0.508***      & 2.508      & 0.475   & 1.104*** & -3.401 & 0.485     & 1.068***   & -2.827   \\
Consumer Cyclicals          & 305   & 0.510    & 1.066***  & 1.689  & 0.206        & 0.628***      & 0.266      & 0.448   & 1.025*** & -3.755 & 0.498     & 1.072***   & -4.060   \\
Consumer Non-Cyclicals      & 205   & 0.483    & 0.829***  & 3.115  & 0.224        & 0.566***      & 1.139      & 0.452   & 0.883*** & -1.971 & 0.472     & 0.885***   & -1.942   \\
Energy                      & 164   & 0.317    & 0.756***  & 4.421  & 0.355        & 0.787***      & -0.231     & 0.549   & 1.195*** & -4.721 & 0.539     & 1.000***   & -2.494   \\
Financials                  & 305   & 0.099    & 0.325***  & 4.362  & 0.132        & 0.641***      & -0.717     & 0.131   & 0.486*** & 0.484  & 0.130     & 0.438***   & 1.378    \\
Healthcare                  & 160   & 0.605    & 1.062***  & 1.662  & 0.312        & 0.701***      & -1.067     & 0.530   & 0.901*** & -2.823 & 0.499     & 0.836***   & -1.950   \\
Industrials                 & 479   & 0.204    & 0.784***  & 3.174  & 0.077        & 0.432***      & 2.390      & 0.337   & 1.053*** & -3.824 & 0.312     & 1.024***   & -3.368   \\
Technology                  & 210   & 0.390    & 0.935***  & 1.994  & 0.159        & 0.509***      & 0.859      & 0.425   & 1.002*** & -3.861 & 0.379     & 0.954***   & -3.261   \\
Telecommunications Services & 91    & 0.590    & 0.869***  & 2.844  & 0.330        & 0.821***      & -1.627     & 0.510   & 0.988*** & -3.536 & 0.517     & 1.038***   & -3.688   \\
Utilities                   & 124   & 0.072    & 0.415**   & 5.195  & 0.089        & 0.587         & 1.096      & 0.227   & 0.990*** & -3.197 & 0.218     & 0.847***   & -1.312   \\
* p\textless{}0.05, **p\textless{}0.01, ***p\textless{}0.001 &       &               &               &                     &                     &             &              &                 &                
\end{tabular}%
}
\end{table}

\subsection*{Water withdrawal}
From the results of the analysis for water withdrawal, it can be concluded that the relationship is weaker, although in most cases still statistically significant for employees, assets and revenue (Table \ref{tab:water}). However, in almost half of the industries, market capitalisation did not show a significant relationship. When looking at the water withdrawal data for the whole sample, it was found that it was best fitted to the regression model of revenue. Yet, the data was best fitted to employees when dividing the sample into industries. Interestingly enough, employees might be a better explanatory variable than total revenue for the water withdrawal.For more than half industries, scaling is superlinear, but, whether an industry scales sub- or superlinear, this feature is generally consistent for employees, assets or revenue as explanatory variables.
\begin{table}[]
\caption {Water withdrawal vs. company size indicator} \label{tab:water} 
\resizebox{\textwidth}{!}{%
\begin{tabular}{llllllllllllll}
                            &       & \multicolumn{3}{l}{Employees} & \multicolumn{3}{l}{Market Capitalisation} & \multicolumn{3}{l}{Assets}  & \multicolumn{3}{l}{Total Revenue} \\
Sector                     & n     & Adj. $\text{R}^2$  & $\beta$   & c      & Adj. $\text{R}^2$      & $\beta$       & c          & Adj. $\text{R}^2$ & $\beta$  & c      & Adj. $\text{R}^2$   & $\beta$    & c        \\
All                         & 2,090 & 0.124    & 0.587***  & 4.119  & 0.061        & 0.449***      & 2.144      & 0.140    & 0.731*** & -0.755 & 0.193     & 0.839***   & -1.570   \\
Basic Materials             & 365   & 0.253    & 0.888***  & 3.885  & 0.145        & 0.616***      & 1.538      & 0.350    & 1.113*** & -3.443 & 0.349     & 1.048***   & -2.578   \\
Consumer Cyclicals          & 222   & 0.487    & 1.094***  & 1.470  & 0.168        & 0.581***      & 0.692      & 0.344   & 0.889*** & -2.483 & 0.322     & 0.873***   & -2.207   \\
Consumer Non-Cyclicals      & 183   & 0.288    & 0.736***  & 3.595  & 0.142        & 0.510***      & 1.786      & 0.294   & 0.810*** & -1.181 & 0.278     & 0.798***   & -1.025   \\
Energy                      & 152   & 0.415    & 0.902***  & 3.683  & 0.311        & 0.735***      & 0.057      & 0.497   & 1.190*** & -4.936 & 0.530      & 1.025***   & -3.025   \\
Financials                  & 257   & 0.070     & 0.289***  & 4.670  & 0.054        & 0.463***      & 1.197      & 0.048   & 0.336*** & 2.201  & 0.062     & 0.329***   & 2.573    \\
Healthcare                  & 154   & 0.607    & 1.149***  & 1.329  & 0.281        & 0.720***      & -1.243     & 0.547   & 0.999*** & -3.754 & 0.548     & 0.968***   & -3.160   \\
Industrials                 & 345   & 0.242    & 0.737***  & 2.886  & 0.113        & 0.441***      & 1.829      & 0.258   & 0.815*** & -1.972 & 0.246     & 0.814***   & -1.824   \\
Technology                  & 182   & 0.316    & 0.851***  & 2.344  & 0.049        & 0.311**       & 2.856      & 0.239   & 0.797*** & -1.863 & 0.235     & 0.786***   & -1.642   \\
Telecommunications Services & 71    & 0.756    & 1.264***  & 0.235  & 0.227        & 0.899***      & -3.282     & 0.484   & 1.266*** & -7.251 & 0.614     & 1.441***   & -8.562   \\
Utilities                   & 169   & 0.088    & 0.745***  & 4.981  & 0.102        & 1.002***      & 1.931      & 0.142   & 1.272*** & -5.117 & 0.154     & 1.222***   & -3.991  \\
* p\textless{}0.05, **p\textless{}0.01, ***p\textless{}0.001 &       &               &               &                     &                     &             &              &                 &                
\end{tabular}%
}
\end{table}

\subsection*{Waste}
The scaling of waste production shows a poor fit, with only a $\text{R}^2$ of 0.14 (Figure \ref{fig3}). The other size variables did not show a better fit, with $\text{R}^2=0.10$ for employees, $\text{R}^2$=0.09 for assets and lastly, $\text{R}^2$=0.04 for market capitalisation. Thus, the analysis for waste showed that the data does not fit the model well at an aggregate level (Table \ref{tab:waste}). The p-values do, however, show that the relationship between all the size variables and waste is statistically significant at the 0.1\% level. Once more the fit improves when splitting the sample per industry. Thus, although waste production does not appear to show universal scaling, it does appear to be industry-specific. Revenue has, on average, the highest adjusted $\text{R}^2= 0.36$,  implying that data fit the regression model with revenue as explanatory variable the best. Likewise Energy Use and Water Withdrawal, Waste production show superlinearity for most sectors and industries.
\begin{table}[]
\caption {Waste production vs. company size indicator} \label{tab:waste} 
\resizebox{\textwidth}{!}{%
\begin{tabular}{llllllllllllll}
                            &       & \multicolumn{3}{l}{Employees} & \multicolumn{3}{l}{Market Capitalisation} & \multicolumn{3}{l}{Assets}  & \multicolumn{3}{l}{Total Revenue} \\
Sector                      & n     & Adj. $\text{R}^2$  & $\beta$   & c      & Adj. $\text{R}^2$      & $\beta$       & c          & Adj. $\text{R}^2$ & $\beta$  & c      & Adj. $\text{R}^2$   & $\beta$    & c        \\
All                         & 1,767 & 0.103    & 0.604***  & 2.200  & 0.035        & 0.375***      & 1.026      & 0.084   & 0.631*** & -1.605 & 0.135     & 0.767***   & -2.733   \\
Basic Materials             & 323   & 0.100    & 0.883***  & 2.301  & 0.077        & 0.680***      & -0.669     & 0.130   & 1.041*** & -4.343 & 0.085     & 0.800***   & -1.834   \\
Consumer Cyclicals          & 182   & 0.241    & 0.793***  & 1.102  & 0.111        & 0.494***      & -0.203     & 0.230   & 0.725*** & -2.602 & 0.293     & 0.832***   & -3.549   \\
Consumer Non-Cyclicals      & 150   & 0.457    & 0.906***  & 0.965  & 0.164        & 0.538***      & -0.389     & 0.381   & 0.943*** & -4.358 & 0.392     & 0.962***   & -4.523   \\
Energy                      & 128   & 0.336    & 0.929***  & 1.421  & 0.209        & 0.643***      & -1.122     & 0.294   & 0.997*** & -5.042 & 0.281     & 0.865***   & -3.528   \\
Financials                  & 183   & 0.003    & 0.119     & 3.395  & 0.012        & 0.313         & 0.783      & -0.004  & 0.068    & 3.098  & 0.002     & 0.134      & 2.541    \\
Healthcare                  & 119   & 0.629    & 1.231***  & -1.027 & 0.277        & 0.719***      & -3.325     & 0.454   & 0.954*** & -5.408 & 0.517     & 0.834***   & -3.950   \\
Industrials                 & 353   & 0.162    & 0.758***  & 1.220  & 0.053        & 0.391***      & 0.712      & 0.197   & 0.902*** & -4.429 & 0.258     & 1.023***   & -5.431   \\
Technology                  & 142   & 0.250    & 0.824**   & 0.412  & 0.061        & 0.378***      & 0.221      & 0.297   & 0.994*** & -5.765 & 0.251     & 0.922***   & -4.982   \\
Telecommunications Services & 60    & 0.559    & 1.184***  & -1.272 & 0.300        & 1.010***      & -6.184     & 0.595   & 1.338*** & -9.807 & 0.566     & 1.344***   & -9.413   \\
Utilities                   & 127   & 0.164    & 0.832**   & 1.889  & 0.063        & 0.599***      & -0.747     & 0.187   & 1.114*** & -6.243 & 0.206     & 1.104***   & -5.638        \\
* p\textless{}0.05, **p\textless{}0.01, ***p\textless{}0.001 &       &               &               &                     &                     &             &              &                 &                
\end{tabular}%
}
\end{table}

\subsection*{Univariate versus multivariate bench-marking}
Corporations are complex systems and reducing the analysis of their metabolism to a bi-dimensional scale is \emph{Procustes' bed}. As shown by the literature for the case of emissions, the more the explanatory variables, the more accurate is the prediction \cite{goldhammer2017estimating,nguyen2021predicting,heurtebize2022corporate}. In the same vein, employing more than one variable to define the size of the corporations it improves the fitting of the scaling. The improvement, tough negligible, is also confirmed by an AIC test (see SM). Nevertheless, the questions is whether the gain in accuracy compensate for the loss in transparency. If the goal of the benchmark is that of enabling a fast and transparent circulation of easily processable information among the stakeholders, a regression based on only one, significant size variable is preferable.  How much accuracy do we loose in availing of only one variables?
For the whole sample of corporations and each sector separately, we performed a multivariate regression analysis with independent variables: Market Capitalisation (MC), Assets (ASS), Revenues (REV) and Employees (EMP). We then compared the observed value of $\text{CO}_{2}e$ emissions with the fitted lines obtained by both the multivariate analysis and the univariate ones (see table 1 in the main text) computing for each one the frequency of points locating above (below)
such regression lines in both cases. In other terms, we compute the number of observations having positive (negative)
$\Delta$multi = $\text{CO}_{2}e$ obs - $\text{CO}_{2}e$ multi and $\Delta$bi = $\text{CO}_{2}e$ obs - $\text{CO}_{2}e$ bi divided by the total number of observations in the sample. Table \ref{tab:compare}reports the share of observed points lying above (below) both the fitting multivariate and univariate
curves. Rows indicate the specific sector; columns report the independent variable (size measure) of the univariate regression. Over all the entire sample, for 93\% of corporations employing revenues as a size factor is enough informative to determine their benchmark in terms of emissions of $\text{CO}_{2}e$.
\begin{table}[]
\caption {Emissons benchmark consistency between univariate and multivariate regression as a share of total corporations.} \label{tab:compare} 
\resizebox{0.5\textwidth}{!}{%
    \centering
    \begin{tabular}{|l|l|l|l|l|}
    \hline
        \textbf{} & \textbf{MC } & \textbf{ASS} & \textbf{REV} & \textbf{EMP} \\ \hline
        \textbf{All} & 0.79 & 0.83 & 0.93 & 0.87 \\ \hline
        \textbf{Basic Materials} & 0.71 & 0.83 & 0.88 & 0.78 \\ \hline
        \textbf{Consumer Cyclicals} & 0.75 & 0.81 & 0.86 & 0.88 \\ \hline
        \textbf{Consumer  Non-Cyclicals} & 0.75 & 0.84 & 0.84 & 0.89 \\ \hline
        \textbf{Energy} & 0.83 & 0.93 & 0.89 & 0.8 \\ \hline
        \textbf{Financials} & 0.85 & 0.92 & 0.96 & 0.91 \\ \hline
        \textbf{Healthcare} & 0.63 & 0.63 & 0.7 & 0.89 \\ \hline
        \textbf{Industrials} & 0.74 & 0.88 & 0.88 & 0.79 \\ \hline
        \textbf{Technology} & 0.73 & 0.83 & 0.83 & 0.83 \\ \hline
        \textbf{Telecommunication Services } & 0.77 & 0.77 & 0.85 & 0.87 \\ \hline
        \textbf{Utilities} & 0.81 & 0.91 & 0.91 & 0.86 \\ \hline
   \end{tabular}%
}
\end{table}
\subsection*{A case study: the insurance \& brokers sector}
\begin{figure}[ht]
\begin{center}
\centerline{\includegraphics[width=1\textwidth]{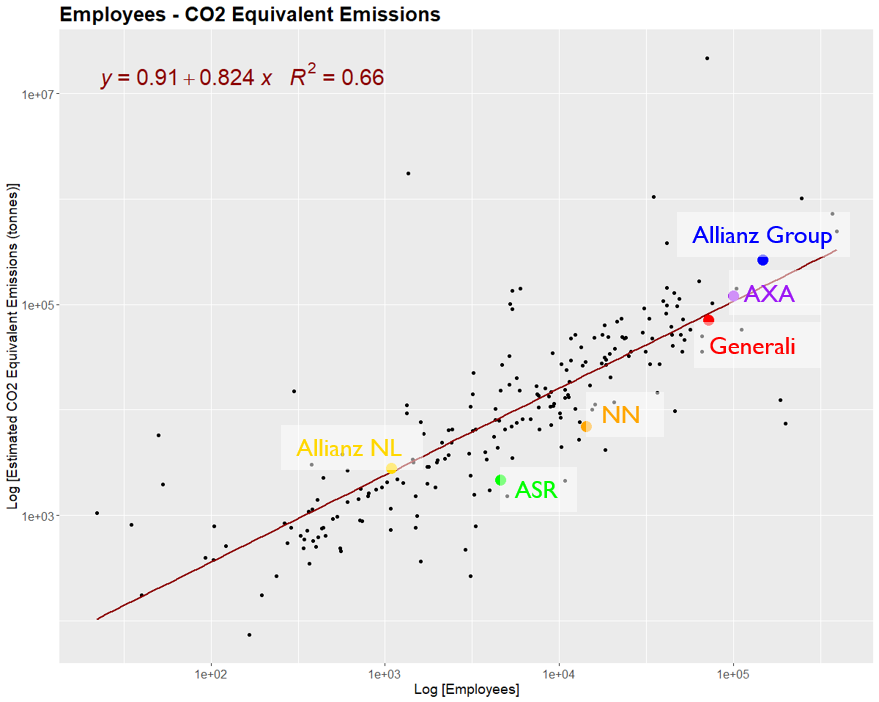}
}
\caption{\textbf{Log-log scale plot of employees vs. estimated $\text{CO}_2$ equivalent emissions of insurance \& brokers TRBC (Thomson Reuters Business Classification) industry. Red line = linear regression, with the regression model depicted on the left top (including the $\text{R}^2$-value) }\label{fig4}}
\end{center}
\end{figure}

In line with the Sectoral Decarbonization Approach, which allows companies to set sector-specific emissions targets according to output intensity indicators \cite{krabbe2015aligning, bjorn2021paris} and in accordance with the theory of Strategic Bench-marking, whose scope stretches from national to global corporations, within and across sectors, as a management practice and investigation methodology \cite{watson2007strategic,walker2021pitfalls}, we herein propose a bench-marking approach based on scaling analysis and self-reporting data. As an example, we consider the case of insurance \& brokers. We compare Allianz, and Allianz NL, with NN Group and ASR Nederland, Assicurazioni Generali and AXA. Employees as an explanatory variable was found to have the highest goodness-of-fit with the data in this sector. Some of their major competitors are highlighted. Figure \ref{fig4} shows that both Allianz Group as AXA are above the benchmark. This indicates that they emit more greenhouse gases than what would be expected on the basis of their size. At the same time, their competitors are below the benchmark, indicating a good environmental performance. Interestingly, however, Allianz has been scoring high in terms of ESG performance in numerous reputation (third-parties) indices, such as, for example: top 5\% in the insurance sector for VIGEO EIRIS; $1^{st}$ at subindustry level for SUSTAINALYTICS; $1^{st}$ in institutional shareholder services for ISS QualityScore;  Top 8\% of sector for FTSE4 GOOD; Gold class (overall best in the sector globally) for DJSI; A+ for PRI and  A- for CDP and AAA for MSCI  \cite{allianz} 
These indices and ratings focus of course on more than only the GHG emissions, but this finding is nevertheless surprising. A possible explanation might be that existing indices and ratings are not weighted for firm size, like scaling does.  Allianz and AXA, for example, are among the two largest insurers in the world, and might therefore be favoured because other ratings are not weighted for firm size.

We also asked ourselves what would be the effect on the environmental impact of our analysis if all companies included in  our analysis would adapt to the existing benchmark. We estimated that using the scaling as a benchmark roughly 15\% of the global emissions could potentially be saved and 11\% of the worldwide energy consumption. In addition, 10\% of the global water withdrawal could be saved. Thus, altogether using scaling in policy could lead to significant reductions in the environmental impact in terms of emissions, energy use, water withdrawal and waste produced.

\section*{Discussion}
That sublinear scaling observed in most industries was in line with expectations, the superlinear cases, however, were not. Superlinearity leads to open-ended growth. Thus, the environmental impact grows at an increasing pace when companies become larger. It is possible that superlinearity might be caused by economies of scale whereby cost savings increase with the level of production. In economy of scales large companies might be able to lower the price of their product or service as the size grow and an higher output would result in an higher environmental footprint. This theory is supported by the fact that the industries in which superlinearity was observed are highly competitive industries, e.g. airlines, oil \& gas, and independent power producers. Another hypothesis points to the role of subsidies for fossil fuels and energy intensive industries, although we must assume these to be universally bestowed. Furthermore, the sub or super linear nature of scaling is important for setting the emissions and environment targets of corporations, because in the former case the efficiency increases with the size, in the latter it decreases, meaning that bigger companies will be either facilitated or hindered in aligning their environmental-climate and economic-financial targets. 
We show the extent of the potential reductions in the environmental impact of companies by enforcing a sector-specific, size-dependent benchmark. However, it was assumed that all companies with values above the benchmark, i.e. companies that have a higher impact than would be expected by their size, can comply with the benchmark value. While it might be theoretically feasible, as other companies are already having an impact below the benchmark, it might not be feasible in practice due to differences in national regulations or local economic and geographical determinants. Some countries have more stringent policies than others, or  more rigid climatic conditions, let alone the large gamut of costs of energy. 
We will refer to these kind of constrains as \emph{the spatial embedding} of corporations (a problem sometimes referred to in the economic literature as that of \emph{international benchmarking} \cite{walker2021pitfalls}).
In order to assess the effect of the spatial embedding on the sample we computed the coefficients of variance (CV’s) for the country of incorporation (the country where the company is legally registered). Most of the countries lie within one standard deviation from the benchmark (see SM2 for more details). In fact, only 11 out of 76 countries are more than one standard deviation away from the benchmark (emissions). Implying that, in general, most countries are relatively close to the benchmark. However, within a country larger standard deviations can be observed. This suggests that in many countries there are companies that perform really good, and other that perform badly. However, the wide standard deviations within countries also indicate that there might not be such structural errors in the countries that either cause an increase or a decrease in the emissions, i.e. no significant spatial embedding (distribution of the CV’s for all the indicators can be found in the Supplementary Material SM2). The problem of interpreting deviations from the fit, or that symmetrical of the actual slope of the $\beta$, is a recurrent in scaling analysis. The study of scaling laws in cities, an other prominent form of social organization which can be considered the closets to our case study within the broad domain of application of scaling laws, showed that deviations can depend on the juridical definition of urban area, the granularity of data or the functional taxonomy \cite{louf2014scaling,bettencourt2010urban} . Seemingly, we may assume that deviations (and a better fitting) can be explained by a more in-depth analysis of featuring peculiarities of each sector/industry/country and on the \emph{system boundaries}, that is, on the administrative contour of a corporation with respect to subsidiaries and parents; besides, obviously, using a moire refined statistical approach to deal with heteroskedasticity and other possible source of bias \cite{bettencourt2010urban,bettencourt2013origins}.

Thirdly, there is the vexed question of the reliability of self-reporting data and ESG rating providers \cite{berg2020rewriting}. The Supplementary Material (SM2) includes an extensive error analysis performed comparing different database and different years of the same database to show that despite some major, localized inconsistencies, our results are generally unaffected by discrepancies in the reporting source. The reason of the soundness of our analysis lies in the volume of data and variety of sources rather than the reliability of the single assessing/reporting provider. It was not detected any systematic error that would significantly hamper our analysis more than any random or sporadic error, fraudulent or fortuitous. 
The success of scaling analysis for organisms built on a theoretical model (the efficient network model) that could explain the mechanism and predict the exponents \cite{west1997general,banavar1999size}. Here, we still far from an fully fledged model, but the fact that for some industries the scaling laws is better using employees rather than revenue as the size variable hints to the fact that when the embedding supply chain is negligible (such in the case of the insurance industry), the underlying network is \emph{social} rather than \emph{economic}; and the metabolism might be determined by the behavior of employees rater than \emph{energy imperatives}, such as the displacement of suppliers, the underlying transport network or the (physical) nature of the featuring economic process.   
\section*{Conclusions}
Scaling has shown to be a promising approach for assessing the environmental impact of companies based on few simple metrics, but conspicuous and available data. The results from the scaling analysis showed that there is a significant relationship (p < 0.001) between the environmental impact variables and size. In other words, changes in a size variable result in changes in the environmental impact variables. A  significant relationship and goodness-of-fit was frequently observed, globally and for every sector. In addition, the results proved to be robust, as the data from previous years showed similar results. Typically the scaling of industries was shown to be sublinear, i.e. the environmental impact of companies increases at a slower pace than the size of the company. Nevertheless, there are also industries that scale superlinearly, which might be possibly due to economies of scale. The implication of these findings for setting environmental benchmarks and targets is twofold: 1) not only the sector, but also the size must be considered in bench-marking; 2) when the sector scale sublinerly the bigger the company the easier will find it to reduce its environmental impact.Finally, to set the benchmark, total revenue should be selected as the explanatory variable, followed by the number of employees and total assets. The notion that is revenue rather than number of employees the main scaling factor of corporate metabolism suggests that the underlying, featuring network (i.e. the network that generates the scaling relationship), is the \emph{supply chain } rather than the \emph{social network of stakeholders}, as suggested by West \cite{west2017scale}, but more research is needed.


It is worth noting that the hereby proposed methodology relies heavily  on a vast and open-access database whereupon single companies or regulators can estimate the current, sector specific benchmark. Ultimately that means that the more data is available and accessible, the more the accuracy of the assessment. In the SM2, we report the needed information to asses the emissions of single companies as compared to the sector's benchmark. It is thus recommendable for international (and, in the future, national/regional) corporation to access to shared information to set their ESG goals more accurately, on the one hand, to enhanced the transparency and accessibility of their data, on the other. Hence, the development of a global and local repository on information in relation to corporate metabolism should be the goal of the policymaker in order to progress toward the much coveted "corporate stewardship" on the path toward sustainability \cite{folke2019transnational}.


\section*{Methods and data \label{sec_methods}}
\subsection*{Data}
This research focuses on publicly traded companies from all over the world. Companies, global operating companies in particular, typically have complex organisational structures including sophisticated supply chains. Thus, while goods might be sold in one country, the environmental impact might be caused in a different country. 

\begin{figure}[ht]
\begin{center}
\centerline{\includegraphics[width=1\textwidth]{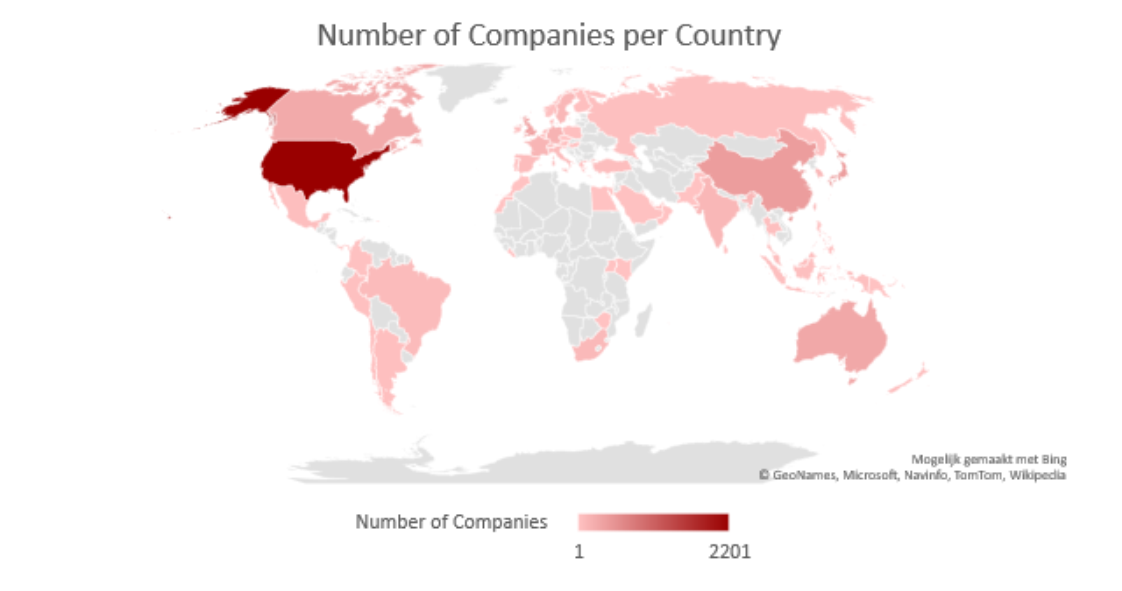}
	}
\caption{\textbf{World map with number of companies per country of incorporation (the legal address of the company and where it pays the corporate taxes) for the estimated $\text{CO}_2$ equivalent emissions sample} \label{fig1}}
\end{center}
\end{figure}

For the analysis Thomson Reuters EIKON (formerly Datastream) is used to collect the data. EIKON is a set of software products provided by Refenitiv for analysing financial markets. It covers 99\% of the global market cap, across more than 150 countries and 133 TRBC (Thomson Reuters Business Classification) industries~\cite{refinitiv2019}. Thomson Reuters acquires information from annual reports, corporate sustainability reports, nongovernmental organisations, and news sources for large, publicly traded companies at annual frequency. It is one of several providers that measure firms’ ESG performance. Differences with other providers, such as Systainalytics and Bloomberg, originate from which ESG choices are considered by each data provider and how they are weighted. 
The data was collected by using the Screener App in EIKON. With this tool you can identify companies that meet certain criteria. The app also allows you, besides the financially related data, to screen on ESG data. In total there are 658 ESG data items, of which 121 on environmental issues.  Figure \ref{fig1} shows the spatial distribution of the reporting companies. 

The data is from 2018 and the initial sample includes 7,587 companies from 81 countries and 133 TRBC (Thomson Reuters Business Classification) industries (Figure \ref{fig1}). However, in order to analyse data, companies with missing or zero values were removed from the dataset and industries with less than 10 companies were excluded. An overview of the samples used for the analyses is shown in Table \ref{tab:indicators} per environmental impact variable analysed. As a comparison, the global GDP in 2018 was 75 trillion € and the annual global $\text{CO}_2$ equivalent emissions were 51.8 gigatonnes. Meaning that the dataset covers 28\% of the global emissions. Likewise, with 104,734 PJ, the Energy Use sample covers 18\% of the global direct primary energy consumption~\cite{ritchie2018}. Compared to the global freshwater use, the Water Withdrawal sample encompasses 11\%.
\begin{table}[]
\caption {Indicators of environmental impact} \label{tab:indicators} 

\resizebox{\textwidth}{!}{%
\begin{tabular}{lllll}
                                        & Estimated $\text{CO}_2$ Eq. & Energy Use & Water Withdrawal & Waste     \\
Total of environmental impact indicator & 14,573 Mt         & 104,734 PJ & 451 billion m3   & 19,068 Mt \\
Number of companies                     & 6,529             & 2,416      & 2,090            & 1,767     \\
Number of Countries                     & 76                & 66         & 63               & 61        \\
Number of Industries                    & 123               & 91         & 79               & 73        \\
Total Number of Employees (x1,000,000)  & 124               & 75         & 68               & 54        \\
Total Revenue (billion €)               & 41,497            & 25,679     & 23,769           & 19,739    \\
Total Assets (billion €)                & 97,774            & 58,476     & 53,611           & 430,176   \\
Total Market Capitalisation (billion €) & 58,785            & 33,209     & 30,126           & 220,175  
\end{tabular}%
}

\end{table}

\subsection*{Environmental Impact indicators, definition}
\begin{description}
     \item  [Estimated $\text{CO}_2$ equivalents emission] The following gases are relevant : carbon dioxide ($\text{CO}_2$), methane (CH4), nitrous oxide (N2O), hydrofluorocarbons (HFCS), perfluorinated compound (PFCS), sulfur hexafluoride (SF6), nitrogen trifluoride (NF3). Total $\text{CO}_2$ emission = direct (scope1) + indirect (scope 2)Scope 1: direct emissions from sources that are owned or controlled by the company Scope 2: indirect emissions from consumption of purchased electricity, heat or steam which occur at the facility where electricity, steam or heat is generated        
     \item [Energy use] The total amount of energy that has been consumed within the boundaries of the company's operations. Total direct and indirect energy consumption: Total energy use is total direct energy consumption plus indirect energy consumption. Purchased energy and produced energy are included in total energy use. For utilities, transmission/ grid loss as part of its business activities is considered as total energy consumed. For utilities, raw materials such as coal, gas or nuclear used in the production of energy are not considered under total energy use. 
     \item [Total water withdrawal] The total volume of water withdrawn from any water source that was either withdrawn directly by the reporting organization or through intermediaries such as water utilities. Different sources of water like well, town/utility/municipal water, river water, surface water, etc. are considered 
     \item [Total amount of waste produced] Total waste is non-hazardous waste plus hazardous waste. Only solid waste is taken into consideration. For sectors like mining, oil \& gas, waste generation like tailings, waste rock, coal and fly ash, etc. are also considered
\end{description}

\subsubsection*{Methods}

Allometry was first introduced in 1936 to describe the discrepancy between the growth rate of body parts in organisms \cite{sullivan2019scaling}. Almost all physiological characteristics scale with body mass (M) according to a power law: $Y= Y_{0}M^{\beta}$, where Y can denote the biological variable (e.g. energy consumption) and the exponent  reflects the general dynamic rule at play~\cite{west1997general}. The research into using scaling as a tool for revealing underlying dynamics and structure has led to a unified quantitative picture of the organisation, structure, and dynamics of organisms. Allometric scaling has been explained by means of network theory whereby organisms were modelled as transportation network (of metabolites and nutrients) maximizing their efficiency \cite{banavar1999size}.
Similar to organisms, social organisations have some kind of metabolism as well. In order to sustain themselves there is a set of flows of materials and energy, suggesting that they might have similar scaling dynamics as organisms.

The scaling laws has also bee successfully applied to cities by Bettencourt, one of the leading scholar on scaling analysis \cite{bettencourt2007, bettencourt2013origins}. They studied if there is quantitative and predictive evidence that support the implications that social organisations are extensions of biology, and proved that this is indeed the case. Companies can also be approached as social system, just like cities; West showed how revenues of companies follow scaling laws, suggesting an underlying (social or economic) network could be the reason \cite{west2017scale,zhang2021scaling}.   The total revenue (or sales) can be thought of as the metabolic trait of the company while the expenses can be thought of as the maintenance costs.  T

Regardless of their industry, all companies cannot produce goods or provide services without creating complex organisational structures. It is essential that these structures are adaptive if it is going to survive in a competitive market. Producing goods or providing services requires the integration of energy, resources and capital – the metabolism of a company. With  $N(t)$ as a measure of the company's size at time $t$, the power law scaling takes the form

\begin{equation}
Y(t)=Y_{0}N(t)^{\beta}
\label{eq0}    
\end{equation}
Where $Y$ denotes the environmental impact variable ($\text{CO}_2$ equivalent emissions, energy use, water withdrawal or waste). $Y_{0}$ is a normalisation constant and the exponent which reflects general dynamic rules across the companies. 
When plotted on a log-log scale, these scaling relationships are linear:
\begin{equation}
\ln{Y(t)}=\beta\ln{N(t)}+\ln{Y_{0}}
\label{eq1}    
\end{equation}
Consequently these relationships can be described using the simple linear equation:
\color{black}
\begin{equation}
y=\beta x + \gamma
\label{eq2}    
\end{equation}
In its linear form, the constants in the scaling equations can be determined by performing a simple regression analysis method, e.g. a least-squares regression which is used in this research. After combining the samples from the different continents, the regression analysis was performed using R Studio (RStudio Team, 2020). For every dependent variable (= the environmental indicator) a linear regression model was fitted using the measures for firm size, i.e. assets, employees, market capitalisation and total revenue, as independent variables. 
The scaling was considered to be superlinear when $\beta$>1.02  and sublinear when$\beta$ <0.98. In the other cases the scaling was considered to be approximately linear between 0.98 and 1.02. A sublinear scaling are typical of a phenomena featured by a stabilizing growth path, which means a growth that tends to a steady state. On the contrary, superlinear features unbounded growth that tends to instability (\cite{west1997general,west2017scale}.

\subsection*{Data availability}
 The data that support the findings of this study are available from Thomson Reuters EIKON  but restrictions apply to the availability of these data, which were used under license for the current study, and so are not publicly available. Data are however available from the authors upon reasonable request and with permission of Thomson Reuters EIKON. All data generated during this study are included in this published article as Supplementary Data.
 
\subsection*{Acknowledgements}
Authors would like to thank Noam Abadi for his reviewing contribution and professor Luis Bettencourt
for his valuable and regarded comments which helped us in improving the paper.
RM acknowledges support from the Italian "Programma di Attività Integrata" (PAI) project "PROsociality COgnition and Peer Effects" (PRO.CO.P.E.), funded by IMT School for Advanced Studies Lucca and the European Union – Horizon 2020 Program under the scheme \lq\lq INFRAIA-01-2018-2019 – Integrating Activities for Advanced Communities\rq\rq{], Grant Agreement n.871042, \lq\lq SoBigData++: European Integrated Infrastructure for Social Mining and Big Data Analytics\rq\rq{} (http://www.sobigdata.eu).

\subsection*{Authors contribution}
RB and RM performed the analysis; RM, YS and FR supervised the analysis; FR designed the analysis. All authors wrote and contributed in reviewing the manuscript.
\bibliographystyle{elsarticle-num} 
\bibliography{bibfile.bib}

\end{document}